\documentclass[]{aa}
\usepackage{graphics}
\begin{document}

\def\oneandonethirdspace{\baselineskip 16pt}

\title{X--ray emission from A0--F6 spectral type stars}

\author{M.R. Panzera\inst{1}
 \and  G. Tagliaferri\inst{1}
 \and L. Pasinetti\inst{2}
 \and E. Antonello\inst{1}
} 

\institute{Osservatorio Astronomico di Brera, via E. Bianchi 46, I--23807
Merate (LC), Italy
\and Dipartimento di Fisica, Universita degli Studi di Milano, Via Celoria 16, 
I-20133 Milano, Italy}

\thesaurus{13.25.5, 08.03.5, 08.03.3, 08.01.2, 08.02.3}

\offprints{M.R.~Panzera}

\date{Received date / accepted date}

\titlerunning{X--ray emission from A0--F6 spectral type stars}
\authorrunning{Panzera et al.\ }
 
\maketitle
                          
\begin{abstract}

We use the ROSAT public data archive to
study the X--ray emission of a sample of supposedly single A0--F6
spectral type stars from the Bright Star Catalogue.
We detected X--ray emission from 19 A-- and 33 F--type stars.
However, our results are not sufficient to associate with certainty
the X--ray emission to the A--type stars themselves, since the usual
argument that it may originate from a binary companion can not be excluded.
A spectral analysis was conducted for 14 sources (3 A and 11 F), finding
that for 12 of
them a two temperature thermal plasma model is needed to reproduce the
observed spectra. The two temperatures are centered at 0.13 and 0.54 keV,
respectively. The values found for the higher temperature are lower than that
ones of X--ray selected single late--type stars. 
The X--ray luminosities are in the range $L_{X} \sim 10^{28} - 10^{30}$
erg s$^{-1}$, with a distribution similar to that of active late--type stars.
No correlation is found between $L_{X}$ and B--V color, 
$V \sin{i}$ and $L_{bol}$, while a positive correlation is found 
between the X--ray luminosity and the hardness ratio.


\end{abstract}

\section{Introduction}

The chromospheric and coronal activity of cool dwarf stars like the Sun
is thought to arise from the interaction of stellar magnetic fields
with differential rotation and convection. 
Since significant convection zones make their appearance along the main 
sequence in late spectral class A, the much debated turn--on of such activity 
should occur at about the same spectral type, although the precise dependence 
is not well known.
On the other hand significant radiatively driven winds 
(the instabilities of which are thought to be responsible
for the X--ray emission in O-- and early B--type stars)
are not present beyond spectral type $\sim$ B4.
Thus in the B5--A5 spectral range a lack of X--ray emission is
theoretically ``expected''.
In spite of the observational efforts, no conclusive evidence of X--ray
emission from B5--A5 type stars has been found; in fact in most of the
observed cases the emission is thought to come not from the early type star
itself, but from a cooler companion or from a nearby star (Schmitt
\& K\"urster 1993; Gagn\'e \& Caillault 1994; Stauffer et al. 1994; Stern
et al. 1995). 
Few interesting cases have been found by Schmitt et al. (1993) and 
Bergh\"ofer \& Schmitt (1994), which detected X-ray emission from both
components in some visual binaries formed by a B5--A5 primary star and
a cooler companion.

Stringent upper limits have been determined for the X--ray emission of some 
well studied A--type stars, down to a luminosity of 
$L_{X} \sim 3.5 \times 10^{25}$ erg\,s$^{-1}$ for the prototypical A--type 
star Vega (Schmitt 1997).
This seems to indicate that the coronae surrounding these stars  (if may
exist at all) must be very different from those surrounding cooler stars. 
They do not have massive winds, neither have deep enough convective zone
for an effective dynamo activity. 
However, in all the cases reported above, only small samples of late B-- early 
A--type stars were used. 
Recently, Simon et al. (1995) tried to overcome this problem by using pointed 
ROSAT PSPC observations to study the X--ray properties of a sample of 74 
A--type stars. 
They detected X--ray emission in 9 late A-- and 10 early A--type stars. 
Of the latter, 5 were confirmed double and 5 were not known to be double but
further optical study are necessary in order to determine if they are really 
single stars.

On the other hand convincing evidence of chromospheric and X--ray emission has
been detected in stars as early as spectral type A7 (Schmitt et al. 1985; 
Simon \& Landsman 1991). Recently Simon \& Landsman (1997) reported the 
detection of chromospheric emission in HST/GHRS spectra of the A4 star 
$\tau^{3}$ Eri that seems to be the hottest main sequence star known to have 
a chromosphere and thus an outer convection zone.
Activity indicators in A7--F5 type stars seem to be independent on rotation, 
while 
the coronal X--ray emission, as measured relatively to chromospheric emission, 
is deficient if compared with later--type stars (Pallavicini et al. 1981; 
Schmitt et al. 1985; Simon \& Landsman 1991). 
The thin convective zone may not maintain strong enough dynamos to support 
{\it solar--like} coronae; rather, it has been suggested that these coronae 
are heated acoustically (Simon \& Landsman 1991; Mullan \& Cheng 1994). 
However, Schrijver (1993) found evidence for a rotation--activity correlation 
in the C II excess flux for these stars, and also the X--ray activity in early 
F stars seems to be coupled with the Rossby number (Schmitt et al. 1985). 
Finally, using the ROSAT All--Sky--Survey, Schmitt (1997) analysed the X--ray 
emission of all A, F and G stars within 13 pc from the Sun and
found that the coronal properties of late A-- early F--type are not
different from those of later type stars in any respect. 
These findings support the presence of a magnetic dynamo as early as 
the A7 spectral type, where convective zones are very shallow.

The ROSAT database provides a powerful instrument for trying to solve 
some of these problems, in particular to verify the existence 
of X--ray emission in early A--type stars and to study the characteristics 
of the X--ray emission in late A-- and early F--type stars. 
Here we present the results obtained for a sample of A0--F6 type stars
obtained by cross-correlating the Bright Star Catalogue (Hoffleit and 
Warren 1991) with the WGA catalogue derived from the ROSAT PSPC public
archive (White, Giommi \& Angelini 1994).

The paper is organized as follows: in Section 2 we present the sample of the
program stars, the X--ray data and their analysis. The basic results are 
given in Section 3, while in Section 4 we discuss our results, compare them
to those obtained for samples of similar and cooler stars.
In Section 5 we present our conclusions.

\section{Observations and Data Analysis}

\subsection{Program Stars}

We selected all stars listed in the Bright Star Catalogue 
(Hoffleit \& Warren 1991) with spectral type A0--F6 (0.0 $<$ B--V $<$ 0.5): 
1963 A-- and 1014 early F--type stars of all luminosity classes.
Since the BSC is complete down to a limiting visual magnitude $V = 6.5$,
the input sample forms a well defined (optically) flux--limited sample
of stars.
This sample was cross--correlated against the WGA catalogue 
of pointed and serendipitous X--ray sources
generated from ROSAT PSPC images (see Tr\"umper (1983) and Pfeffermann
et al. (1986) for a description of the ROSAT satellite).
We accepted all sources whose coordinates were inside the one arcmin 
region radius of the optical coordinates.
Of the 351 A-- and 165 F--type stars that could have been observed in our
X-ray images, 142 were detected; 66 A-- and 76 F--type stars.
We tested the probability that one of our optical candidate 
falls inside the one arcmin error box by chance. We shifted the optical
coordinates of our sample of A--F stars by 30 arcmin and then
cross-correlated with the WGA catalogue.
We found that the probability of mis--identification is very small, 
of the order of 0.3~\%.
Nevertheless we searched for all the plausible optical counterparts present in the 
142 X--ray error boxes and rejected all cases in which the emission could be 
attributed either to the A0--F6 star or to another optical source on the
basis of its 
X--ray--to--optical flux ratio (e.g. Stocke et al. 1991; Giommi et al. 1991). 
This search was done using the Space Telescope Guide Star Catalogue
(Lasker et al. 1990) and SIMBAD database (we could not use the digitized
sky survey plates because these stars are so bright that the plates are
totally saturated around these regions). In this step 5 A--
and 3 F--type stars were rejected.

Since our purpose is to study the X--ray emission from supposedly
{\it single} A-- and early F--type stars, all confirmed or {\bf suspected}
binaries or multiple systems not resolved by the PSPC were rejected
(42 A-- and 40 F--type stars), regardless of the nature of the companion.
For instance we rejected HD\,61421 (Procyon), although
for this nearby astrometric binary, consisting
of a slightly evolved F5 IV-V star and a cool white dwarf companion,
the observed X--ray emission is attributed to the F--type star
(see Schmitt et al. 1985). In any case the coronal activity of the
system is already well studied (see e.g. Lemen et al. 1989, Schmitt
et al. 1990 and Schmitt et al. 1996). Another star that we rejected
is HD\,203280 (Alderamin, A7IV), that is listed as an X-ray source
by various authors (e.g. Simon et al. 1995; Schmitt 1985). However,
the SIMBAD database lists a second star at $\sim 15 \arcsec$
(ADS 14858 CD, V=10.4), thus following our strict criteria we excluded
this source.
We note that we would reject also the binary systems formed by two
early A--type stars (or by an early A-- and late--B star). These stars
are not expected to be X-ray emitters, so their detection would be of
extreme interest (see also Berg\"ofer et al. 1996). Schmitt
et al. (1993) and Bergh\"{o}fer \& Schmitt (1994), using the ROSAT HRI
detector to study visual binary systems formed by a B--type star
with a cool companion, found that in some cases also the B star was 
an X--ray source. However, again we would not be able to determine
with our data which star would be responsible for the X-ray emission.
Moreover of the 42 rejected A--type stars, only one is a
visual binary system formed by an early A and a late B--type star
separated by 8\arcsec (A1V + B9V, HD\,24071).

The WGA catalogue has not been manually cleaned and/or inspected
field by field. By re--analysing interactively all images, one
could find some more stars lost by the WGA catalogue.
These stars could either be slightly below the WGA detection limit
or be in ``difficult'' fields (crowded or with a very bright source),
or lie near the ribs or the border of the detector.
Moreover, the latest version of the WGA catalogue has been generated
from the PSPC observations available in the public archive at
HEASARC in February 1995 for a total of 3007 pointings. The good PSPC
observations available at the HEASARC are now 4597. We checked how many
sources we are probably loosing in the 1590 PSPC fields missing from
the WGA. There are 180 pointings (114 for A-- and 66 for F--type stars)
in which 36 new A-- and 28 new F--type stars could
have been detected. If we apply the detection rate that we found for
the WGA fields, we should have missed about $\sim 7$ and $\sim 13$ stars for 
the A-- and F--type, respectively. If we consider that most of them would
be in binary systems, we are really missing very few stars.
Thus, the stars that we are presenting here
are not ALL the single A-- and F--type stars possibly detected in PSPC
pointings. But this does not affect our goals, since we are not
deriving a X--ray luminosity function. Our aim is a) to look for supposedly
single A stars that are X-ray emitters, and b) to study the characteristics
of the X--ray emission of single A and F stars. Besides, sources are
randomly missed from the WGA and the results should not be biased. 

The final sample, here presented, consists of 52 supposedly
single stars, {\bf 19} A-- and {\bf 33} early F--type stars, associated 
with X--ray sources in {\bf 74} ROSAT PSPC fields
(29 fields for the A--type 
stars and 45 fields for the early F type stars). 
The sample contains 30 main--sequence stars, 9 sub--giants, 1 giant, 2
super--giants, 6 stars with uncertain spectral type and/or luminosity class
and 4 stars without any information about the luminosity class;
there are 5 chemically peculiar stars and three $\delta$ Scuti stars
(HD\,124953, HD\,432 and HD\,89449).
In Table 1 we list the 52 stars by their HD number (Column 1). 
Columns 2, 3, 4 and 5 contain the spectral type, magnitude,
B--V color and distance of the stars in parsec.
Except for HD\,124953, all these informations were taken 
from the HIPPARCOS Output Catalogue (Perryman et al. 1997).
For HD\,124953 the informations are from Garc\'\i a et al. (1995).
Note that the HIPPARCOS distances often are different from those
previously known or derived from spectroscopic parallaxes.
This could imply a refinement of the spectroscopic classification
(few sub-classes) for various stars. However, given that there is not
a systematic effect, this should not have implications on our results
as a sample.
Column 6 lists the projected rotational velocity  
from the Bright Star Catalogue and from Uesugi \& Fukuda (1982).
Column 7 contains the bolometric luminosity  
calculated following Ayres et al. (1981) and  
column 8 lists the ROSAT sequence number of the images.
The A--type stars already studied in the literature have been marked
with a note after the HD number and the relevant references
are reported at the table bottom.

\subsection{X--ray observations and Data Analysis}

\subsubsection{Count rates}

The count rates listed in the WGA catalogue present some limitations for 
our study due to: 
(1) the automatic detection technique used, based on an optimized sliding cell 
algorithm has problems in the case of crowded fields, extended sources,
etc.;
(2) the count rates are obtained using a constant value of the exposure time 
(nominal time) across the field, thus 
the source count rate can be underestimated if the source lies close to the 
detector ribs or the border; 
(3) the band used to obtain the count rates corresponds to channels 
24--200 or 0.24--2.0 keV, this excludes the band 0.1--0.24 keV, which 
is extremely important for stellar X--ray sources. 
Thus the WGA catalogue was used only to identify the sources detected 
by the PSPC from our optical sample. 
Each observations was 
then re--analyzed interactively using the XIMAGE package.
Count rates were measured in the energy band $0.1 - 2.4$ keV 
(PSPC channels 11--235) by processing the event files retrieved from the
ROSAT public archive. The effective exposure times were measured from the 
exposure maps provided by the ROSAT Standard Analysis Software System (SASS). 
These exposure maps account for the telescope vignetting, the occultation
effects due to the support ribs in the detector window and the ``wobble" of 
the spacecraft. 
These corrections are often important for our X--ray sources because 
almost all of them have been detected serendipitously, i.e. they were 
in the field of view of other targets ($\sim 45\%$ of 
our sources were located near the detector ribs or border).
We also compared our results with those obtained from the ROSAT All--Sky
Survey (RASS--BSC, Voges et al. 1996; RASS, H\"{u}nsch et al. 1998). 
Of our supposedly single stars, 8 A-- and 26 F--type stars are detected
also in the RASS. We found good agreement in the count rate values implying
that no strong variability is present.

In order to obtain the usual hardness ratio 
HR=$(H - S)/(H + S)$ (see e.g. Schmitt et al. 1995), we
calculated the count rates in the ``soft" ({\it S}: channels 11--41 
$\approx 0.1 - 0.28$ keV) and ``hard" ({\it H}: channels 52--201 
$\approx 0.5 - 2.0$ keV) PSPC X--ray bands, respectively.
A value of HR $\simeq -1$ indicates an extremely soft spectrum,   
HR $\simeq 0$ indicates an equal number of soft and hard photons
and HR $\simeq +1$ an extremely hard one.
The results of our data analysis are also tabulated in Table 1,
column 9 lists the PSPC count rate and
column 10 lists the hardness ratio.

\subsubsection{Spectral Analysis}

The PSPC has a moderate spectral capability, with an energy resolution 
$\Delta E / E \approx 0.42$ at 1 keV and, in principle, it allows
to study the temperature of the plasma in the 0.1--2.4 keV energy band. 
However, as it is typical for serendipitous detections, the majority of our 
sources lacks the required signal--to--noise ratio to model the spectral energy 
distribution. Only for few of them, 3 A-- and 11 early F--type stars, 
there are enough counts to perform a detailed spectral analysis. For
this analysis we considered only the sources with more than 800 counts.
The X--ray pulse height spectra were extracted from the event files
using the XSELECT/FTOOLS package and were rebinned to give at least 25 
counts per bin. Standard Poissonian statistics should therefore be fairly 
adequate to compute the errors in the model fitting. The spectral analysis
was carried out using the optically thin plasma model ``mekal" (see Mewe et
al. 1985; Mewe et al. 1986; Kaastra 1992), as implemented in the XSPEC package.
The hydrogen column density $N_{\rm H}$ was added in the spectral fit. 

\section{results}

\subsection{Spectral Results}

For all sources, we compared the best fit values of the 
$N_{\rm H}$ with those obtained using Paresce's formula (Paresce 1984). 
Note that all these stars have d\,$ < 50$ pc, but one at 65 and one at 96 pc.
For 12 out of 14 sources, a two--temperature model (with solar 
abundances) reproduces observations better than a single temperature model. 
In 5 of the 12 cases (HD\,50241, HD\,116160, HD\,25457, HD\,45348, HD\,89449) 
a further improvement was obtained by introducing the $N_{\rm H}$ as a free 
parameter.
For 4 of them the value of the $N_{\rm H}$ obtained from Paresce's 
formula is a factor 2--25 smaller than the lower limit of the $N_{\rm H}$
derived from the best fit (90\% confidence range for three parameters of
interest: the two temperatures and the $N_{\rm H}$).
For the fifth source (HD\,45348) the 90\% confidence range is 
consistent with the value obtained from Paresce's formula 
(see also Bauer \& Bregman 1996).
For the remaining two cases nor the second temperature neither the inclusion 
of the $N_{\rm H}$ does improve the fit: these are HD\,13456 (F5V) and 
HD\,18404 (F5IV). In this analysis we used an F--test to check if the 
improvement of the $\chi^2$ was significant at more than 99.75\%.
 
In order to characterize the X--ray spectra of these sources we used
the 2T mekal model with solar abundances for all sources.
Due to the low energy resolution of the PSPC and to the relatively low 
number of photons of our spectra, in Table 2 we report the 
errors at the 90\% confidence level only for the two temperatures.
The $N_{\rm H}$ was fixed to Paresce's value for the 9 sources for 
which its inclusion did not improve the fit, and to the best--fit 
value in the remaining 5 cases.

\begin{table}
\caption[]{Spectral Results}
\begin{tabular}{l l l l l l}
\hline
&&&&&\\

$HD$  &$kT_{1}$ &$Norm_{1}$  &$kT_{2}$  &$Norm_{2}$  
&$N_{\rm H}$($\times 10^{19}$)  \\
&(keV)    &($\times 10^{-4}$) &(keV)     &($\times 10^{-4}$)  
&(cm$^{-2}$) \\

&&&&&\\
\hline
&&&&&\\

50241   &$0.09^{+0.02}_{-0.03}$  &5.27  &$0.57^{+0.13}_{-0.15}$  &0.95  &20.5 \\
116160  &$0.10^{+0.03}_{-0.03}$  &7.49  &$0.69^{+0.11}_{-0.10}$  &5.10  &14.6 \\
187642  &$0.11^{+0.02}_{-0.03}$  &2.85  &$0.42^{+0.35}_{-0.19}$  &0.48  &0.10 \\
13456   &$0.17^{+0.0}_{-0.09}$   &0.47  &$0.47^{+0.43}_{-0.40}$  &1.00  &1.00 \\
18404   &$0.13^{+0.06}_{-0.05}$  &0.69  &$0.43^{+0.20}_{-0.0}$   &1.38  &0.64 \\
20675   &$0.14^{+0.04}_{-0.05}$  &0.35  &$0.53^{+0.16}_{-0.15}$  &0.57  &0.92 \\
24357   &$0.18^{+0.71}_{-0.08}$  &0.38  &$0.56^{+0.33}_{-0.29}$  &0.47  &0.83 \\
25457   &$0.08^{+0.01}_{-0.01}$  &64.5  &$0.58^{+0.04}_{-0.04}$  &27.3  &17.1 \\
37495   &$0.17^{+0.07}_{-0.08}$  &0.90  &$0.64^{+0.23}_{-0.19}$  &1.50  &0.85 \\
40136   &$0.10^{+0.04}_{-0.02}$  &2.38  &$0.34^{+0.09}_{-0.04}$  &3.92  &0.30 \\
45348   &$0.14^{+0.03}_{-0.04}$  &7.09  &$0.68^{+0.10}_{-0.09}$  &10.1  &11.2 \\
89449   &$0.09^{+0.01}_{-0.02}$  &4.94  &$0.53^{+0.06}_{-0.08}$  &2.09  &16.3 \\
124850  &$0.17^{+0.05}_{-0.06}$  &3.87  &$0.66^{+0.09}_{-0.08}$  &12.5  &0.43 \\
197373  &$0.14^{+0.04}_{-0.04}$  &0.81  &$0.51^{+0.21}_{-0.14}$  &0.96  &0.67 \\
\hline
&&\\

\end{tabular}
\end{table}

We note that, a 2T model is required also for the 3 A--type stars 
in our sample: Altair (HD\,187642, A7IV-V), HD\,50241 (A7IV) and HD\,116160 
(A2V). 
Altair is an unquestionably single star and a known X--ray and chromospheric 
source (Schmitt et al. 1985; Simon et al. 1995; Blanco et al. 1980; 
Landsman \& Simon 1991; Simon et al. 1994; Walter et al. 1995; 
Simon \& Landsman 1997).
Up to now, the changeover from radiative to convective envelopes 
is presumed to occur in the close vicinity of its B--V color (see Simon \& 
Landsman 1997).
This star has a soft X--ray spectra with the majority of X--ray 
photons recorded in the lowest energy channels below 0.5 keV and a hardness 
ratio HR=$-$0.8. 
However, unlike previous reported {\it Einstein} and ROSAT results 
(Golub et al. 1983; Schmitt et al. 1990; Freire Ferrero et al. 1995; 
Simon et al. 1995), we
find that the inclusion of a second temperature improves the fit. 
The first temperature (0.11 keV) is consistent with that obtained in 
the single temperature fit  of the previous works, while the 
second temperature has a value of 0.42 keV and it contributes 
to about 20\% of the total flux.
HD\,50241 and HD\,116160 are supposedly single A--type stars, although
the few informations found in the literature do not exclude a cooler binary 
companion. If HD\,50241 is really a single star, this will be a further
confirmation that coronal activity can already be present
in stars as early as spectral type A7.
In case of HD\,116160, the spectral type (A2V) and the high X--ray 
luminosity detected ($\log{L_{x}} \sim 30$; see also Simon et al. 1995) 
could imply that this star has a cool companion.
The distributions of $T_{low}$ and $T_{high}$ for the
14 sources are shown in Fig. 1 (solid line), and are
discussed in section 4.

The spectra of four F--type star, out of the 11 presented here, 
are already studied in literature using {\it Einstein} and/or 
ROSAT observations (HD\,37495, HD\,40136, HD\,45348 and  HD\,124850).
For HD\,124850 our results are well consistent with those obtained 
by Maggio et al. (1994) analysing the same image. Instead Bauer \& 
Bregman (1996) fit the X--ray spectra of HD\,40136, HD\,45348 and HD\,124850
with a single--temperature model and free metal abundance.
They found a good fit with very low metal abundance values ($\sim 0.1$
of the solar value) and a temperature that is similar to our harder
temperature.
Finally the X-ray spectra of HD\,37495 and HD\,45348 obtained with the
{\it Einstein} satellite were fitted using a single temperature model
by Schmitt et al. (1990); the resulting temperature are much harder
($\sim 1.3$ keV) and this is probably due to the harder band of
the {\it Einstein} satellite (0.3--3.5 keV).

\vskip 0.5truecm

In order to give a crude spectral information also for the other
sources of the sample (38 objects), in column 10 of Table 1 we report 
the hardness ratio (HR) for all sources.
An inspection of the HR values shows that they are almost all negative 
(61 out 74), clustering around the value HR=$-$0.2. 
The value of HR immediately gives a rough idea of the coronal temperature 
distribution for a sample of stars, if the interstellar absorption is 
unimportant. 
In our case the majority of stars is closer than 60 pc, thus the 
absorption should be negligible. 
Actually, the source with the hardest spectrum (HD\,20902, F5Ib with HR=0.7) 
is also the farthest (d=$181 \pm 22$ pc) and in this case the absorption 
could be important: Paresce's formula gives a value of 
N$_{\rm H} \sim (4 \pm 0.4) \times 10^{19}$
cm$^{-2}$ which is too small to explain the observed HR. However, this formula
is not reliable for such a distance and a value of 
N$_{\rm H} \sim 5 \times 10^{20}$
cm$^{-2}$ would already be sufficient to absorb all the soft photons.

\subsection{X--ray Flux and Luminosity}

To estimate the integrated energy flux, it is necessary to multiply the count 
rate by a conversion factor (CF).
In principle, the CF depends upon the effective area of the instrument as a 
function of energy, which is known, and the incident X--ray spectrum of the
source that, for most of our sources, can not be determined because there are
not enough counts to carry out a spectral analysis.
Thus, the conversion from the measured count rate to the energy
flux requires some assumptions for the intrinsic source spectrum.
On the other hand, as the PSPC is sensitive over a large energy range, 
we can try to construct an empirical relation between the
HR (indicator of the spectral shape in case of small interstellar
absorption) and the CF of the stars of our sample for which we performed the
spectral analysis.
This relation would allow to obtain a CF from the HR value for 
the other stars of the sample. 
This attempt was already done for late--type stars samples with a wide 
range of stellar activity levels and spectral shapes 
(Fleming et al. 1995; Panarella et al. 1996). 
However, in our case, we were not able to obtain a relation between HR and CF.
This is probably due to the fact that our stars with sufficient counts 
to conduct spectral analysis (14 stars) have a too small range 
of HR ($-0.3 <$ HR $< 0.2$ except for Altair)
to be compared with the range between $-1$ and $+1$ of the late--type 
star samples. In any case, if we consider all our sample we find that 
in 51 out of 74 observations the 
A--F stars have $-0.5 <$ HR $< 0$ (see Table 1, column 10).
Thus, for our stars the HR has little influence over the CF. 
Therefore we decided to use a constant value, CF$ = 5 \times 10^{-12}$ 
erg cm$^{-2}$ cts$^{-1}$, obtained from the mean of the 14 CF of the 
stars with enough counts for spectral analysis.

We note that all distances, but one, are taken from the HIPPARCOS
Output Catalogue (Perryman et al. 1997) which provides extremely 
precise measurement of the parallax and of the other astrometric 
parameters (1 milliarcsec level astrometry). Thus, the X--ray 
luminosity is independent of the measurement of quantities
such as spectral type and color index.
In column 11 of Table 1 we list the derived luminosities in erg s$^{-1}$ 
(in the 0.1--2.4 keV energy band).
The X--ray luminosity distribution is shown in Fig. 2, panel a).

\section{Discussion}

\subsection{X--ray Temperatures}

As in our spectral sample we have only 3 A--type stars, our
results are dominated by the early F stars.
We find that a thin thermal plasma model with two temperature 
components agrees with those obtained for late--type stars 
with {\it Einstein},
EXOSAT, Ginga, ROSAT and ASCA observatories (e.g. Schmitt et al. 1990; 
Dempsey et al. 1993; White 1995; Panarella et al. 1996).
In Fig. 1 we show the distributions of the two--temperature 
components (solid line) for the 14 sources of our sample. From this figure
and from Table 2 it emerges that $T_{low}$ is in the range between 0.06 and 
0.2 keV, while $T_{high}$ is in the range between 
0.3 and 0.7 keV, clustering around values of 0.13 keV 
and 0.54 keV, respectively.
In order to compare our temperatures with those of late--type stars,
in the same figure we report the distributions of the two
temperatures for a X--ray selected sample of late--type
stars (dashed line).
The sample consists of EXOSAT serendipitous sources (Giommi et al. 1991,
Tagliaferri et al. 1994) for which a spectral analysis like
ours was conducted using  ROSAT PSPC data (Panarella et al. 1996).
Also for this sample a two temperature model was required to represent 
the spectra of all sources. 
For this comparison we considered only the supposedly single 
late--type stars of the sample (15). 
In this way we compare the temperatures of active supposedly single A--F type 
stars with those of active supposedly single G--K--M type stars.

The comparison of the temperature distributions shows that, in our
case, the second temperature is significantly lower. In fact for late--type stars
the second temperature is in the range 0.5 -- 1.2 keV with
an average value of $\sim$ 0.9 keV while for A--F type stars the average
value is around 0.54 keV. A Kolmogorov--Smirnov test gives a probability
of $6.8 \times 10^{-5}$ that the two high temperature distributions are
coming from the same parent distribution.

\begin{figure}
\begin{center}
\resizebox{\hsize}{!}{\includegraphics{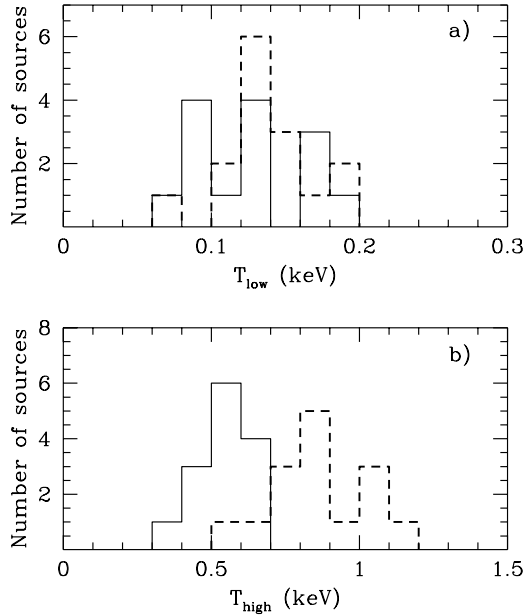}}
\end{center}
\caption[]{Distribution of $T_{low}$ (panel a) and $T_{high}$
(panel b) components for the 14 sources of our sample with enough counts for
spectral analysis (solid line).
In order to compare our temperatures with those of late--type stars we show 
also the distribution of the two temperatures for a sample of single late--type 
stars (dashed line). See discussion in the text.}
\end{figure}

\subsection{X--ray Luminosity}

In Fig. 2, panel a) we plot the distribution of the derived ROSAT 
luminosity $L_{X}$ for our sample. From this figure and from Table 1 
(column 11) it is clear that: 
(1) the X--ray luminosity distributions of A-- and early F--type 
stars are similar; 
(2) the X--ray luminosities range from  $L_{X} \sim 10^{28}$ erg s$^{-1}$ 
up to $L_{X} \sim 10^{30}$ erg s$^{-1}$, levels 
similar to those observed in active late--type stars;
(3) X--ray luminosities cluster around the same value of  
$<\log L_{X}> = 29.2$ for A-- and early F--type stars.

In Fig. 2 we compare the X--ray luminosity distribution
of our sample to that of two sample of late--type stars.
Panel b) shows the X--ray luminosity distribution of all known K and M dwarfs 
in the immediate solar vicinity with distance less than 7 pc 
(Schmitt et al. 1995). 
In this study the X--ray detection rate for K and M dwarfs is 87 \%.
Thus, this sample should be reasonably complete, except possibly for the 
very faintest stars, and it is not biased toward the intrinsically luminous 
emitters like the X--ray selected samples of coronal X--ray sources.
Inspection of panel b) shows that the luminosity 
ranges from $L_{X} \simeq 10^{26}$ erg s$^{-1}$ to 
$L_{X} \simeq 10^{29}$ erg s$^{-1}$ clustering
in the range $L_{X} = 1-3 \times 10^{27}$ erg s$^{-1}$.
Panel c) shows the luminosity distribution for the 
sample of the supposedly single late-type stars 
(G--K--M spectral type stars) studied by Panarella et al. 1996. 
As this is flux limited X--ray selected sample, it consists
of the most active G--K--M stars. 
Their luminosity is spread in the 
range $L_{X} \simeq 10^{28} - 10^{30}$ erg s$^{-1}$.

The comparison of panels a), b) and c) shows that our luminosity 
distribution is similar to that of the X--ray selected sample of late--type 
stars, but not to that of late--type optically selected stars.
We tried to characterize the X--ray emission of our A-- and early 
F--type stars searching for correlations between $L_{X}$ and the B--V 
color, $V \sin{i}$ and the bolometric luminosity, finding none. 
On the contrary a positive correlation is found between $\log{L_{X}}$
and the spectral hardness ratio. A linear regression analysis 
yields a correlation coefficient $r = 0.6$, the probability of
a spurious correlation being less than $10^{-10}$.
A similar trend was found for samples of late--type stars  
(see Schmitt et al. 1995; Schmitt 1997; Panarella et al. 1996).

\begin{figure}
\begin{center}
\resizebox{\hsize}{!}{\includegraphics{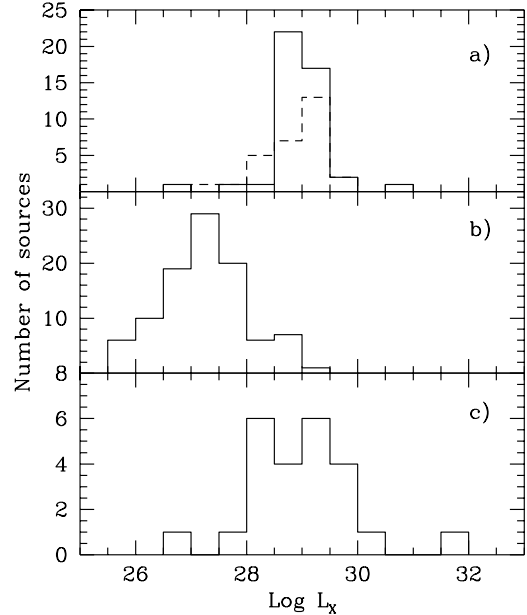}}
\end{center}
\caption[]{In panel a) we plot the distribution of the derived ROSAT X--ray 
luminosity for A-- (dashed line) and early F--type stars 
(solid line) of our sample.
In order to compare our distribution to that of late--type stars, in panels
b) and c) we show the X--ray luminosity distributions for a sample of
``normal'' and very active late--type stars, respectively. 
See the text for a description of the samples and discussion.}
\end{figure}

\subsection{Comparison with previous Observations of A--type stars}

The detection of a very stringent upper limit for the X--ray
emission of the prototypical A--star Vega 
($\log{L_{X}} < 25.55$; Schmitt et al. 1997), which is well below our values,
raises doubts about the existence of coronae in early A--type stars.
In a work similar to our, Simon et al. (1995) used ROSAT PSPC
observations to study a sample of 74 A--type stars detecting X--ray
emission in 10 early A--type stars. Of these five are known to be
binaries, while more optical observations are necessary to determine
the physical nature of the remaining five (four of them are also in our
sample: HD20888, HD30478, HD45618 and HD116160).
In the past X--ray emission was reported for the Ap stars as a class.
The main results on these stars can now be summarized as follows.
Pre--ROSAT studies of chemically peculiar stars (CP) reported only a small 
number of X--ray detection of putatively single CP stars (see Cash \& Snow 
1982; Golub et al. 1983).
Drake et al. (1994) searched for X--ray emission in $\sim$ 100 magnetic
Bp--Ap stars using the ROSAT All Sky Survey database. 
They detected X-ray emission for ten of these sources, but only 
three of them were identified with apparently single stars. Thus,
these stars do not seem to be X--ray emitters as a class
any more. 
A wind--fed magnetosphere model has been proposed to explain both the 
non--thermal radio and the X--ray emission found in some CP stars 
(Linsky et al. 1992).
Cash \& Snow (1982), using {\it Einstein} observations, did not detect 
single Am stars and suggested that Am stars as a class are not strong coronal 
emitters (upper limit of about $10^{28}$ erg s$^{-1}$). 
One apparently single Am stars (HD\,107168, A8m) was detected 
by Randich et al. (1996) using ROSAT PSPC observations. 
They suggested a cool companion as source of the emission
since the star lies slightly above cluster main sequence.
We also detected one magnetic Ap star (HD\,38104) and two Am
stars (HD\,140232 and HD\,187753).
HD\,38104 is given as an Ursa Major stream star (Roman 1949).
Schmitt et al. (1990) studied this region using {\it Einstein} 
observations, finding for this star only an upper 
limit $\log{L_{X}} < 28.78$ (with a distance
d=52 pc). We instead find a value $\log{L_{X}} = 29.1$, using the
HIPPARCOS distance d=148 pc. With this distance the $3 \sigma$ 
upper limit of {\it Einstein} is fully consistent with our detection.
Note that the higher HIPPARCOS distance does not imply that the star is not
an Ursa Major stream, which include stars at similar or higher distance
(e.g. Roman 1949, Eggen, 1958). The two Am stars were detected also in the 
ROSAT all-sky survey at a similar flux level by H\"unsch et al. (1998)
(see also the RASS -- BSC revision 1RXS, Voges et al. 1996).
In our sample we have another peculiar star: HD\,124953, a variable $\delta$ 
Scuti star (L\'{o}pez de Coca et al. 1990; Garc\'\i a et al. 1995) with 
$L_{X} = 2 \times 10^{28}$ ergs s$^{-1}$.
It belongs to the Ursa Major stream and, as HD\,38104, it was not 
detected by {\it Einstein} (Schmitt et al. (1990), with 
an upper limit $L_{X} < 4 \times 10^{28}$ erg s$^{-1}$,
a value again consistent with our detection.
HD\,124953 has been studied also by Ayres et al. (1991), using the 
same image analysed in our work, they found $L_{X} = 1 \times 10^{28}$ 
ergs s$^{-1}$.

\section{Conclusions}

We studied the X--ray emission for a sample of A-- and
early F--type stars which are not known to be double, using the data in the 
ROSAT PSPC public archive. Starting from a list of 351 A-- and 165 F--type
stars that could have been detected in our images, X-ray emission 
was detected in 66 A-- and 76 F--type stars. The rejection of all
confirmed or suspected binaries gives a final sample of  19 A-- and
33 F-- supposedly single type stars.
As expected, the detection rate is much higher for the F--type stars,
given that stars earlier than A7 spectral type are not expected to have
strong (if at all present) X-ray emission.

The main conclusions of this paper can be summarized as follows:

\noindent
1. 19 supposedly single A--type stars are found to coincide 
with X--ray sources.
In many cases we found very little information in literature about
these stars. With the only exception of Altair, our results are 
not sufficient to associate with certainty the X--ray emission to 
the A--type stars themselves, since the usual argument that 
it may originate from a binary companion can not be excluded.
Optical and UV studies are needed to establish the physical properties of 
the early A--type stars detected in X--ray. 
For this reason we observed in the optical a sub--sample of our
A--type stars, monitoring their radial velocities.
The reduction of these data is in progress.

\noindent
2. A two temperature thermal plasma model is a good representation
of the spectra of the 11 F-- and 3 A--type stars with enough counts
to model the spectral energy distribution.
The temperatures found for the 3 A-- stars are consistent with those
of the F--type stars.
An interesting case is that of HD\,116160, an early spectral type star
(A2V), with a high X--ray luminosity.
The comparison of the two temperatures found for the stars in our sample 
with those of a X--ray selected 
sample of single late--type stars (Panarella et al. 1996) shows that, while  
T$_{low}$ is similar for the two samples our second temperature
component (T$_{high}$) is lower. This result applies mainly to the
F-type stars, because our spectral study is dominated by
them (11 out of 14).

\noindent
3. The X--ray luminosity distribution of our sample is similar to that
of X--ray selected sample of late--type stars. 
We find a positive correlation between the X-ray luminosity and the HR,
while we do not find significant correlations between X--ray luminosity and 
other stellar parameters.

\acknowledgements{This research was financially supported by Italian Space
Agency and the Italian Consorzio Nazionale di Astronomia e Astrofisica.
This research has made use of the archival database 
maintained at GODDARD-NASA by HEASARC, as well as of the SIMBAD database,
operated at CDS, Strasbourg, France and HIPPARCOS database. 
We thank Lorella Angelini and Mike Corcoran for their help
with the ROSAT data analysis. We also thank the referee, J.H.M.M. Schmitt,
for his helpful comments which help us in improving an earlier version
of this paper.

\begin{table*}
\caption[]{A0--F6 type stars}
\begin{tabular}{l l l l l l l l l l l}
\hline
&&&&&&&&&&\\
$HD$   &$ST$  &$V$   &$B$--$V$ &$d$ &$V \sin{i}$  &$\log L_{bol}$ &Image &Count Rate   &HR  &$\log L_X$ \\
       &      &      &         &(pc) &(km s$^{-1}$) &(erg s$^{-1}$) &    &($10^{-3}$ cts/s)  & &(erg s$^{-1}$) \\     
&&&&&&&&&&\\
\hline
&&&&&&&&&&\\
16555$^1$    &A6V   &5.30  &0.29   &$44 \pm 4$   &239    &34.63   &RP700103N00 &$18.7  \pm 3.0$  &$ -0.63 $  &28.34 \\
16555        &...   &...   &...    &...          &...    &...     &RP300201N00 &$25.5  \pm 1.8$  &$ -0.33 $  &28.47 \\
16555        &...   &...   &...    &...          &...    &...     &RP701356N00 &$18.9  \pm 1.2$  &$ -0.59 $  &28.34 \\
20888$^2$    &A3V   &6.03  &0.13   &$58 \pm 2$   &...    &34.56   &RP150010N00 &$26.3  \pm 2.8$  &$~~0.13$  &28.72 \\
20888        &...   &...   &...    &...          &...    &...     &RP200044N00 &$51.5  \pm 4.9$  &$ -0.04 $  &29.02 \\
20888        &...   &...   &...    &...          &...    &...     &RP600045N00 &$37.1  \pm 2.6$  &$ -0.11 $  &28.87 \\ 
20888        &...   &...   &...    &...          &...    &...     &RP600504N00 &$32.4  \pm 2.3$  &$ -0.23 $  &28.81 \\
30321        &A2V   &6.33  &0.04   &$85 \pm 8$   &...    &34.78   &RP200997N00 &$13.7  \pm 1.9$  &$ -0.15 $  &28.77 \\
30478$^3$    &A8/A9III-IV &5.28  &0.20   &$68 \pm 2$   &177    &35.00   &RP701021N00 &$22.7  \pm 1.9$  &$ -0.32 $  &28.80 \\
38104$^4$    &A0p   &5.46  &0.03   &$148 \pm 19$ &21     &35.59   &RP900487N00 &$9.7   \pm 1.4$  &$~~0.40 $  &29.10 \\
39014$^5$    &A7V   &4.34  &0.22   &$44 \pm 1$   &206    &35.00   &WP141851N00 &$6.3   \pm 1.4$  &$ -0.04 $  &27.86 \\
45618$^6$    &A5V   &6.61  &0.18   &$69 \pm 3$   &55     &34.46   &WP201603N00 &$9.4   \pm 0.7$  &$ -0.10  $ &28.43 \\
45638$^7$    &A9IV  &6.59  &0.29   &$61 \pm 4$   &42     &34.34   &RP900355N00 &$104.2 \pm 7.5$  &$ -0.24  $  &29.37 \\
45638        &...   &...   &...    &...          &...    &...     &RP900355A01 &$115.0 \pm 10$   &$ -0.41  $  &29.41 \\
45638        &...   &...   &...    &...          &...    &...     &RP900355A02 &$90.0  \pm 7.0$  &$ -0.30  $  &29.30 \\
50241$^8$    &A7IV  &3.24  &0.23   &$30 \pm 2$   &205    &35.11   &WP200638N00 &$92.0  \pm 3.2$  &$ -0.31 $  &28.70 \\
82380        &A4V   &6.76  &0.12   &$91 \pm 7$   &155    &34.69   &RP400244N00 &$29.4  \pm 6.3$  &$ -0.30  $  &29.16 \\
107966$^{9}$  &A3V   &5.17  &0.08   &$87 \pm 5$   &54     &35.28   &WP200307N00 &$20.7  \pm 2.6$  &$~~0.13$    &28.97 \\
111893       &A7V   &6.30  &0.17   &$112 \pm 9$  &215    &35.02   &WP800393A01 &$17.4  \pm 2.5$  &$ -0.20  $  &29.12 \\
116160$^{10}$  &A2V   &5.69  &0.05   &$65 \pm 3$   &200    &34.82   &RP300322N00 &$331.8 \pm 10$   &$ -0.04 $  &29.92 \\
124953$^{11}$  &A9V   &5.98  &0.28   &49           &125    &34.46   &RP150018N00 &$13.1  \pm 1.3$  &$ -1.00  $  &28.27 \\
140232$^{12}$  &A2m   &5.80  &0.21   &$53 \pm 2$   &68     &34.61   &RP700893N00 &$106.9 \pm 5.1$  &$ -0.01 $  &29.25 \\
159312$^{13}$  &A0V   &6.48  &6E-03  &$104 \pm 10$ &...    &34.95   &RP200098N00 &$103.4 \pm 14$   &$~~0.10 $  &29.83 \\
175938$^{14}$  &A8V   &6.40  &0.29   &$88 \pm 4$   &113    &34.81   &RP700058N00 &$40.5  \pm 2.3$  &$ -0.05 $  &29.27 \\
175938       &...   &...   &...    &...          &...    &...     &RP700412N00 &$42.3  \pm 7.2$  &$ -0.04 $  &29.29 \\
175938       &...   &...   &...    &...          &...    &...     &RP700951N00 &$48.9  \pm 5.6$  &$ -0.13 $  &29.36 \\
175938       &...   &...   &...    &...          &...    &...     &RP700950N00 &$57.9  \pm 6.3$  &$~~0.13$  &29.43 \\
187642$^{15}$  &A7IV-V   &0.76  &0.22   &$5 \pm 0$ &242    &34.57   &WP200898N00 &$194.6 \pm 3.8$  &$ -0.81 $  &27.46 \\
187753$^{16}$  &A1m   &6.25  &0.10   &$117 \pm 15$ &10     &35.11   &WP200898N00 &$38.01 \pm 3.9$  &$~~0.30 $  &29.49 \\
432     &F2III-IV &2.28  &0.38   &$16.7 \pm 0.2$  &70  &35.01  &WP201520N00  &$246.7 \pm 10$   &$ -0.53 $  &28.61 \\
8723    &F2V      &5.35  &0.39   &$26 \pm 1$      &61  &34.14  &WP701220N00  &$265.1 \pm 11$   &$ -0.19 $  &29.03 \\
13456   &F5V      &6.00  &0.42   &$50 \pm 2$      &10  &34.48  &RP800114N00  &$87.9  \pm 3.4$  &$ -0.09 $  &29.12 \\
18404   &F5IV     &5.80  &0.42   &$32 \pm 1$      &28  &34.16  &WP900138N00  &$102.0 \pm 4.5$  &$ -0.27 $  &28.80 \\
20675   &F6V      &5.94  &0.47   &$46 \pm 1$      &... &34.44  &WP200223N00  &$39.4  \pm 6.1$  &$ -0.15 $  &28.70 \\
20675   &...      &...   &...    &...             &... &...    &WP200239N00  &$53.8  \pm 7.5$  &$-0.10 $  &28.83 \\ 
20675   &...      &...   &...    &...             &... &...    &WP200228N00  &$48.6  \pm 13$   &$~~0.02$  &28.79 \\
20675   &...      &...   &...    &...             &... &...    &WP200228A01  &$65.6  \pm 15$   &$ -0.17 $  &28.92 \\
20675   &...      &...   &...    &...             &... &...    &WP200222A01  &$49.5  \pm 13$   &$ -0.40  $  &28.80 \\
20675   &...      &...   &...    &...             &... &...    &WP200222N00  &$29.3  \pm 11$   &$ -0.20  $  &28.57 \\
20675   &...      &...   &...    &...             &... &...    &RP201510N00  &$49.3  \pm 1.7$  &$ -0.19 $  &28.80 \\
&&\\
\hline
\end{tabular}
\end{table*}

\begin{table*}
\addtocounter{table}{-1}

\caption[]{{\it (Continue\/)}}
\begin{tabular}{l l l l l l l l l l l} 
\hline
&&&&&&&&&&\\
$HD$   &$ST$  &$V$   &$B$--$V$ &$d$  &$V \sin{i}$  &$\log L_{bol}$ &Image &Count Rate   &HR  &$\log L_X$ \\
       &      &      &         &(pc)  &(km s$^{-1}$) & (erg s$^{-1}$) &    &($10^{-3}$ cts/s)  & &(erg s$^{-1}$) \\     
&&&&&&&&&&\\
\hline
&&\\

20902   &F5Ib     &1.79  &0.48   &$181 \pm 22$    &18   &37.27  &RP201468N00  &$18.9  \pm 1.3$  &$~~0.70 $  &29.57 \\
24357   &F4V      &5.97  &0.35   &$41 \pm 2$      &59   &34.31  &RP200107N00  &$62.2  \pm 2.2$  &$ -0.17 $  &28.80 \\
25457   &F5V      &5.38  &0.52   &$19.2 \pm 0.3$  &23   &33.90  &WP900140N00  &$1548  \pm 21$   &$ -0.09 $  &29.53 \\
28556   &F0V      &5.40  &0.26   &$46 \pm 2$      &95   &34.63  &RP200945N00  &$98.5  \pm 5.7$  &$ -0.20 $  &29.10  \\
28736   &F5V      &6.37  &0.42   &$43 \pm 2$      &35   &34.19  &RP700063N00  &$103.4 \pm 8.8$  &$ -0.42 $  &29.06 \\
28736   &...      &...   &...    &...             &...  &...    &RP700945N00  &$102.0 \pm 8.2$  &$ -0.15 $  &29.05 \\
28736   &...      &...   &...    &...             &...  &...    &RP700916N00  &$90.3  \pm 6.2$  &$ -0.28 $  &29.0 \\
28736   &...      &...   &...    &...             &...  &...    &RP700913N00  &$94.5  \pm 7.8$  &$ -0.14 $  &29.02 \\
28736   &...      &...   &...    &...             &...  &...    &RP700919N00  &$86.1  \pm 7.7$  &$ -0.37 $  &28.98 \\
29169   &F5IV     &6.01  &0.38   &$44 \pm 1$      &80   &34.32  &WP200677N00  &$64.3  \pm 16$   &$ -0.60  $  &28.87 \\
31362   &F0       &6.33  &0.35   &$43 \pm 2$      &60   &34.19  &RP300178M01  &$62.6  \pm 5.0$  &$ -0.09 $  &28.84 \\
37495   &F5V      &5.28  &0.49   &$42 \pm 1$      &31   &34.61  &WP701094N00  &$180.7 \pm 6.9$  &$ -0.12 $  &29.28 \\
38393   &F7V      &3.60  &0.48   &$9 \pm 0.1$     &11   &33.96  &WP200643N00  &$18.9  \pm 3.3$  &$ -0.90  $  &26.96 \\
40136   &F1V      &3.71  &0.34   &$15 \pm 0.2$    &...  &34.34  &RP200907N00  &$292.4 \pm 6.9$  &$ -0.34 $  &28.60 \\
45348   &F0Ib     &$ -0.62$  &0.16   &$96 \pm 5$  &15   &37.70  &WP200319N00 &$586  \pm 15$   &$~~0.21$     &30.51 \\
48737   &F5IV     &3.35  &0.44   &$18 \pm 0.3$    &70   &34.62  &RP201487N00  &$445.8 \pm 18$   &$~~0.08$  &28.94 \\
71243   &F5III    &4.05  &0.41   &$19 \pm 0.2$    &36   &34.43  &WP300131N00  &$271.4 \pm 13$   &$ -0.16 $  &28.77 \\
87141   &F5V      &5.71  &0.51   &$47 \pm 2$      &10   &34.54  &WP700264N00  &$96.5  \pm 7.4$  &$ -0.19 $  &29.11 \\
89449   &F6IV     &4.78  &0.45   &$21 \pm 0.4$    &18   &34.22  &RP200076N00  &$164.2 \pm 3.1$  &$ -0.14 $  &28.64 \\
91480   &F1V      &5.16  &0.35   &$26 \pm 0.4$    &87   &34.25  &WP201310N00  &$93.2  \pm 5.4$  &$ -0.13 $  &28.58 \\
95310  &Am/F0Vs   &5.06  &0.26   &$124 \pm 10$    &72   &35.60  &RP200943N00  &$23.9  \pm 1.7$  &$~~0.08$  &29.34 \\
101688  &F2IV-V   &6.65  &0.36   &$61 \pm 3$      &60   &34.431  &WP701149N00  &$72.6  \pm 6.1$  &$ -0.10 $  &29.21 \\
117361  &F0IV     &6.42  &0.40   &$78 \pm 4$      &...  &34.70   &RP800047N00  &$31.1  \pm 3.2$  &$ -0.60 $  &29.05 \\
118646  &F3V      &5.81  &0.43   &$49 \pm 2$      &...  &34.54   &WP600188N00  &$91.5  \pm 9.3$  &$ -0.06 $  &29.12 \\
118646  &...      &...   &...    &...             &...  &...     &WP600188A02  &$108.9 \pm 3.1$  &$~~0.02$  &29.19 \\
124850  &F7V      &4.07  &0.51   &$21 \pm 0.4$    &17   &34.52   &RP200908N00  &$803.9 \pm 17$   &$~~0.15$  &29.33 \\
125451  &F5IV     &5.41  &0.39   &$26 \pm 1$      &42   &34.14   &WP150053N00  &$199.1 \pm 8.0$  &$ -0.26 $  &28.91 \\
125451  &...      &...   &...    &...             &...  &...     &RP200543N00  &$259.4 \pm 16$   &$ -0.15 $  &29.02 \\
145100  &F3V      &6.43  &0.45   &$49 \pm 2$      &...  &34.31   &RP200545N00  &$18.8  \pm 3.6$  &$ -0.50 $  &28.43 \\
148048  &F5V      &4.95  &0.39   &$30 \pm 0.4$    &76   &34.46   &WP141829N00  &$144.1 \pm 9.0$  &$ -0.14 $  &28.89 \\
155203  &F3III-IVp &3.32 &0.44   &$22 \pm 0.4$    &150  &34.83   &RP200132N00  &$145.5 \pm 8.6$  &$ -0.69 $  &28.62 \\
157373  &F6V      &6.36  &0.43   &$40 \pm 1$      &15   &34.13   &WP701080N00  &$9.8   \pm 1.1$  &$ -0.42 $  &27.97 \\
186185  &F5V      &5.49  &0.46   &$37 \pm 3$      &21   &34.41   &RP600148N00  &$166.7 \pm 9.0$  &$ -0.22 $  &29.14 \\
197373  &F6IV     &6.02  &0.44   &$33 \pm 1$      &30   &34.12   &RP600272N00  &$112.0 \pm 3.5$  &$ -0.33 $  &28.86 \\
&&\\
\hline
\\
\end{tabular}
\small
notes: $^1$ a, b; $^2$ a, b; $^3$ a; $^4$ c; $^5$ b; $^6$ a; $^7$ g;
$^8$ a, g; $^{9}$ d; $^{10}$ a, g; $^{11}$ c, e; $^{12}$ g; 
$^{13}$ g; $^{14}$ a, b, g, h, i, l, m, n; $^{15}$ a, b, g, o, p, q;
$^{16}$ g; a: Simon et al. 1995; b: Schmitt et al. 1985; c: Schmitt,
Micela et al. 1990; d: Randich et al. 1996; e: Ayres et al. 1991; 
f: Simon and Drake 1993; g: H\"unsch et al. 1998; h: 
Helfand and Caillault 1982; i: Gioia et al. 1990; 
l: Stocke et al. 1991; m: Fleming et al. 1995; 
n: Pan et al. 1997; o: Schmitt 1997; p: 
Schmitt et al. 1990; q: Golub et al. 1983.

\end{table*}

\end{document}